# Magnetic Yoking and Tunable Interactions in FePt-Based Hard/Soft Bilayers

Dustin A. Gilbert,[1,2] Jung-Wei Liao,[3] Brian J. Kirby,[2] Michael Winklhofer,[4,5,6]
Chih-Huang Lai,[3] and Kai Liu[1,*]

[1]*Dept. of Physics, University of California, Davis, California 95616, USA*
[2]*NIST Center for Neutron Research, Gaithersburg, Maryland 20899, USA*
[3]*Dept. of Materials Science and Engineering, National Tsing Hua University, Hsinchu 30013, Taiwan*
[4]*Dept. of Earth and Environmental Sciences, Geophysics, Munich University, 80333 Germany*
[5]*Faculty of Physics, University of Duisburg-Essen, 47057 Duisburg, Germany*
[6]*IBU, School of Mathematics and Science, Carl von Ossietzky University, 26129, Oldenburg, Germany*

**Abstract**

Assessing and controlling magnetic interactions in magnetic nanostructures are critical to nanomagnetic and spintronic explorations, such as magnetic recording media, permanent magnets, magnetic memory and logic devices, etc. Here we demonstrate an extremely sensitive magnetic yoking effect and tunable interactions in FePt based hard/soft bilayers mediated by the soft layer. Below the exchange length, a thin soft layer strongly exchange couples to the perpendicular moments of the hard layer; above the exchange length, just a few nanometers thicker, the soft layer moments turn in-plane and act to yoke the dipolar fields from the adjacent hard layer perpendicular domains. The evolution from exchange to dipolar-dominated interactions is experimentally captured by first-order reversal curves, the $\Delta M$ method, and polarized neutron reflectometry, and confirmed by micromagnetic simulations. These findings demonstrate an effective yoking approach to design and control magnetic interactions in wide varieties of magnetic nanostructures and devices.

*Correspondence and requests for materials should be addressed to K.L. (email: kailiu@ucdavis.edu).

Assessing and controlling interactions in magnetic nanostructures are critical to nanomagnetic and spintronic explorations, such as ultrahigh density magnetic recording,[1,2] exchange spring composites for permanent magnets,[3] artificially structured model systems of spin ice[4-6] and analog memory,[7] and magnetic quantum-dot cellular automata,[8-10] etc. For example, in the emerging heat-assisted magnetic recording (HAMR) technology that separates the recording process at elevated temperatures from the ambient temperature storage environment,[11,12] a critical issue is to tailor the various magnetic interactions in the media.[13-15] These interactions are often complicated and correlated, e.g., exchange interaction within each grain vs. dipolar interactions across neighboring grains, which directly impact the media performance such as the thermal stability and switching field distribution. Careful balance of these interactions, e.g., through the introduction of boron or oxides at grain boundaries,[16,17] has been essential to optimizing the media performance. Another example is the exchange coupled media that uses a magnetically hard layer to anchor the thermal stability, while writeablity is achieved by exchange coupling to a neighboring magnetically soft layer.[18-20] The same concept has also been used to achieve high energy density permanent magnets, through enhancement of the maximum energy product.[3] Careful balancing of exchange and dipolar interactions are again important, as the physical dimensions of the magnetically hard/soft components need to be on the order of exchange length, to achieve these desirable effects.

In this work we demonstrate an extremely sensitive magnetic yoking effect and continuous tuning of interactions in a model system consisting of a magnetically hard $L1_0$-FePt layer with perpendicular magnetic anisotropy (PMA), a prototype HAMR media, and a magnetically soft layer with a varying thickness. The latter is shown to effectively moderate interactions between adjacent domains in the hard layer. By tailoring the soft layer thickness by just a few nanometers, across the exchange length, the dominant interaction in the bilayer system can be continuously tuned from exchange to dipolar. A yoking effect emerges as the soft layer magnetization turns in-plane to facilitate the dipolar interaction across neighboring $L1_0$-FePt domains with perpendicular magnetization, captured by magnetometry, polarized neutron reflectometry (PNR), magnetic imaging, and confirmed by micromagnetic simulations. These results demonstrate an effective yoking approach to tailor interactions in nanomagnetic building blocks for wide varieties of technological applications.

**RESULTS**

**Magnetometry.** Thin film bilayers of $L1_0$-Fe$_{52}$Pt$_{48}$ (4 nm)/$A1$-Fe$_{52}$Pt$_{48}$ ($t_{A1}$) with $t_{A1}$ = 0-9 nm were fabricated by sputtering, as described in Methods. Magnetometry measurements were performed at room temperature in the out-of-plane geometry, unless otherwise noted. Hysteresis loops for the films are shown in Fig. 1. For the 4 nm $L1_0$-FePt film alone ($t_{A1}$ = 0 nm), a square hysteresis loop is observed with a coercivity of 320 mT in the perpendicular geometry (panel a), while the in-plane loop is closed (panel b), indicating a clear perpendicular anisotropy expected of (001) oriented $L1_0$-FePt. For films with increasing $t_{A1}$, in the perpendicular geometry the coercivity decreases and the loop develops a strong canting; magnetic moments associated with the slope are due to the in-plane soft layer, which is being reversibly forced out-of-plane by the field. Note that for the sample with $t_{A1}$ = 2 nm, the hysteresis loops are very similar to those for the $L1_0$-FePt alone, suggesting that the $A1$ layer orientation is dominated by the $L1_0$ layer through exchange coupling; this is consistent with the $A1$-layer's exchange length of $l_{ex}$=3.9 nm.[21] Once $t_{A1}$ exceeds $l_{ex}$, significant in-plane magnetization is observed for $t_{A1}$ = 5 nm and even more so for $t_{A1}$ = 9 nm (Fig. 1), indicating more and more of the moments in the $A1$ layer are now in the film plane. The development of this in-plane reversal indicates that the $A1$ layer is able to switch with only limited dependence on the perpendicular $L1_0$ layer.

**Polarized neutron reflectometry**. The interfacial coupling was directly observed using specular PNR which is only sensitive to in-plane moments. Model fitted reflectometry data for $t_{A1} = 2$ nm and 5 nm bilayers, measured in a 50 mT field applied perpendicular to the film, after sequential saturation in the out-of-plane and then in-plane directions, are shown in Figs. 2a and 2b, respectively. The sample with a $t_{A1} = 2$ nm, panel a, shows no appreciable signal in the spin-flip channel ($R_{+-}$ and $R_{-+}$), indicating negligible in-plane magnetization. In contrast, the $t_{A1} = 5$ nm sample, panel b, exhibits spin-flip scattering with intensity of the same order of magnitude as the non-spin-flip, identifying a significant in-plane magnetization. This is further confirmed in the fitted models (Figs. 2c, 2d), which show in-plane magnetization ($\rho_M$) only in the 5 nm sample. These PNR results are consistent with the magnetometry measurements shown in Fig. 1: when $t_{A1} < l_{ex}$, the magnetocrystalline anisotropy in the $L1_0$-FePt forces the moments in both layers to be out-of-plane; however, this effect is largely suppressed once $t_{A1} > l_{ex}$, resulting in significant in-plane magnetization. The non-spin-flip scattering depends only on the depth-dependent nuclear composition, and is consistent with the expected layer structure.

**First-order reversal curves and $\Delta M$ method**. Details of the magnetization reversal in $L1_0$-FePt (4 nm) /$A1$-FePt($t_{A1}$) were identified using the first-order reversal curve (FORC) technique,[22-28] as shown in Fig. 3. The FORC diagrams all show two key features: (1) a horizontal ridge, parallel to the $H$ axis, and (2) a vertical ridge, essentially parallel to the $H_R$ axis. These type of features are commonly seen in PMA films which reverse by a domain nucleation/growth mechanism.[23,29,30] For the $L1_0$-FePt film alone, the horizontal FORC ridge is aligned to the left of the vertical ridge (as in a flipped L-shape), Fig. 3a, noted as "left bending" hereafter. As $t_{A1}$ increases to 2 nm, Fig. 3b, the FORC features remain qualitatively the same, but shift to more positive values of $H_R$ and more negative values of $H$. With further increase in $t_{A1}$ (now $> l_{ex}$), the horizontal feature moves to the right of the vertical one, noted as "right-bending" hereafter (Figs. 3c and 3d). While both left- and right-bending features have been observed in the literature,[23,29-34] there has been no distinction between them nor discussion of the origin of their left/right facing orientation.

We suggest that the orientation of the FORC features indicates the nature of the interactions, particularly between the $L1_0$ domains; the reversible behavior of the in-plane $A1$ layer measured in the out-of-plane direction, identified in Fig. 1a by the high field canting, are manifested in the FORC distribution as a reversible feature at $H=H_R$ (not shown). To qualitatively identify the interactions in the $L1_0$-FePt (4 nm) /$A1$-FePt($t_{A1}$) films we employ the $\Delta M$ technique (see Methods).[35-37] At $t_{A1} = 0$ nm and 2 nm, whose FORC distributions show the left-bending characteristics, the $\Delta M$ plot shows a prominent positive feature, indicating primarily an exchange dominated interaction (Fig. 3e). At $t_{A1} = 5$ nm and 9 nm, whose FORC features are right-bending, the $\Delta M$ plot exhibits a negative feature (Fig. 3e), confirming a change in the dominant interaction from exchange to dipolar.

**Magnetic force microscopy**. The relative balance between exchange and dipolar interactions in these films can also be seen directly in the remanent domain structures captured by magnetic force microscopy (MFM) after perpendicular ac demagnetization, as shown in Fig. 4. At $t_{A1} = 0$ nm, the domain structure of $L1_0$-FePt film consists of large labyrinth domains (average size $8\times10^4$ nm$^2$, Fig. 4a). With increasing $t_{A1}$, the average domain size is shown to decrease ($7\times10^4$ nm$^2$, $3\times10^4$ nm$^2$, $2\times10^4$ nm$^2$ average for $t_{A1}$ = 2 nm, 5 nm, and 9 nm, shown in Figs. 4b-d, respectively). In a simplistic consideration, for a PMA film, domain size is an indicator of the balance between the lateral exchange coupling and the dipolar interaction.[38] Generally, the domain wall energy per unit area increases with decreasing domain size, thus a strong exchange interaction favors fewer domain walls and larger domains; the demagnetization energy decreases with decreasing domain size, therefore a strong dipolar interaction favors smaller domains. In the present case, the magnetically hard $L1_0$-FePt film remains the same in all samples; as the soft $A$1 layer gets thicker, the rapid decrease in domain size suggests a relative increase of dipolar interactions, over exchange interactions, in the bilayer film.

**DISCUSSION**

The $\Delta M$ measurements and MFM results suggest that the left- and right-bending FORC features are correlated with an exchange and dipolar-dominated interaction, respectively. Following directly from Davies *et al.*,[23,29] along decreasing reversal field $H_R$, the horizontal FORC ridge in Figs. 3a-3d marks the nucleation of reversal domains in the early stage of the magnetization reversal from positive saturation, and the vertical ridge (the range of $H_R$) marks the later stage of annihilation of residual domains towards negative saturation. The applied field $H$-position of the vertical ridge also marks the nucleation field of reversal domains from those negative reversal fields, towards positive saturation. Indeed, with increased $A$1 layer thickness this nucleation field occurs sooner (vertical feature moves to -$H$). This is consistent with increased dipolar interactions, which will promote early reversal domain nucleation. Similar results are also found in the $L1_0$-FePt/Fe samples (see Supplementary Figs. S1-S3 online). For samples with $t_{A1} < l_{ex}$, the $A1$ layer is strongly exchange coupled to the $L1_0$ layer, characteristic of exchange spring magnets,[18,39] with the magnetic moments aligned out-of-plane and the hard layer coercivity reduced. In the FORC distribution the coercivity reduction is manifested by the translation of the FORC feature to more positive values of $H_R$ and more negative values of $H$. When $t_{A1} > l_{ex}$, the major hysteresis loop and PNR results indicate an in-plane magnetic easy axis for the soft $A1$ layer. In the perpendicular $L1_0$ layer, neighboring domains will have exactly antiparallel alignment due to its uniaxial anisotropy (as shown in Fig. 4), thus leading to strong dipole fields. The in-plane component of the dipolar field is estimated to be much larger than the soft layer coercivity (see Supplementary Information online). Thus, the $A1$ layer will have in-plane magnetizations which follow the dipolar fields from the $L1_0$ layer and form magnetic yokes to channel the flux.[40] The resultant dipolar interaction between adjacent domains in the $L1_0$ layer will be enhanced by the additional moments in the yoke, as observed in the FORCs, $\Delta M$ and MFM measurements.

The results from our micromagnetic simulations underpin the yoking mechanism inferred from the experimental data. As long as $t_{A1} < l_{ex}$ (2 nm in Fig. 5a), yoking is practically absent; magnetization in the soft phase is strongly exchange coupled to that in the hard phase, which results in a narrow Néel-type domain wall (nearly 180 degrees) with only a minor change in domain-wall width when going from the hard to the soft layer. The out-of-plane and in-plane component of the wall magnetization closely follow the theoretical curve given by $\tanh(x/\sqrt{A/K_u})$ and $\text{sech}(x/\sqrt{A/K_u})$, respectively (dashed curves). For a $t_{A1} \sim 2l_{ex}$ (Fig. 5b), the domain wall in the upper part of the soft phase widens considerably and has a smaller amplitude variations due to the now predominant in-plane magnetization component in the soft layer. From the energy diagram (Fig. 5c), it can be seen that the demagnetizing energy for $t_{A1} = 9$ nm is greatly reduced compared to $t_{A1} = 2$ nm. This gain is a direct consequence of flux closure due to yoking and more than offsets the cost in exchange energy needed to deflect the soft-layer magnetization in-plane. For the case that the soft-layer domain is orthogonal to the hard-layer domain wall, shown in Fig. 5d for the modeled strip (1000 nm long, 10 nm wide) and Fig. 5e for the large block, the soft layer follows the dipolar fields forming a yoke structure, as described above. Figure 5d also demonstrates a multi-domain reversal in the thick $A1$ layer which is not commensurate with the domain structure in the $L1_0$ underlayer. Interestingly, despite their incommensurate domain structures, for the case that the soft-layer domain is oriented parallel to the domain wall, Fig. 5f, the yoking effect is still present and re-orients the soft-layer near the hard-layer domain wall to follow the dipolar fields. Thus, independent of the domain structure in the soft-layer, yoking is expected to occur and facilitates the dipole interactions.

In summary, a magnetic yoking effect is observed in $L1_0$-FePt based hard/soft bilayers with perpendicular magnetization. The soft layer is shown to play a critical role in continuously tuning interactions between perpendicular magnetic moments in adjacent hard layer domains. When the soft layer is thinner than the exchange length, its moments strongly exchange couple to the $L1_0$-FePt and remain perpendicular to the film; as the soft layer becomes thicker than the exchange length, its moments turn in-plane and act to yoke the dipolar fields between neighboring hard-layer domains. This extreme sensitivity to the soft layer thickness, over a few nanometers, and the corresponding transition from exchange to dipolar dominated interactions are captured by $\Delta M$ measurements, as well as PNR and domain size analysis, and confirmed by micromagnetic modeling. These results suggest orientation of the boomerang FORC distribution may identify the dominant magnetic interactions. The yoking approach presents an effective mechanism to tailor magnetic interactions in elemental nanomagnetic building blocks, which have far-reaching potentials in numerous technological applications such as magnetic recording media, permanent magnets, magnetic memory and logic devices, etc.

## METHODS

**Synthesis.** Thin films of $Fe_{52}Pt_{48}$ (4 nm) were grown by dc magnetron co-sputtering from elemental targets on Si substrates with $SiO_2$ (200 nm)/$FeO_x$ (<1 monolayer) buffer layers in a high-vacuum chamber ($P_{Base}$ < 1×10$^{-4}$ Pa, $P_{Ar}$ = 3×10$^{-1}$ Pa). After deposition of the FePt, the film was exposed to a short oxygen plasma treatment (4:1 Ar:$O_2$), realizing a thin oxide capping layer. The films were subsequently heated by rapid thermal annealing to 500° C for 5 min in vacuum, following prior procedures.[41-43] X-ray diffraction indicates that the resultant films are $L1_0$ ordered with an (001) texture (not shown), similar to previous studies.[42] Thus this magnetically hard $L1_0$-FePt has its easy axis perpendicular to the film. Next, an $A1$-$Fe_{52}Pt_{48}$ film (with thickness $t_{A1}$ = 0-9 nm) and a 2 nm Ti capping layer were dc magnetron sputtered, which is magnetically soft with an easy axis in the film plane. The bilayer structure thus forms an exchange spring system.[18-20] Additional bilayer films with an Fe soft layer, rather than an $A1$-FePt, were also fabricated, as discussed in the Supplementary Information online.

**Characterizations.** Magnetic properties were measured by a vibrating sample magnetometer (VSM) at room temperature with the applied field perpendicular to the film, unless otherwise stated. Details of the magnetization reversal were investigated using the FORC technique, which can identify different reversal mechanisms and distinguish interactions.[22-25,27,28] Starting at positive saturation the applied field, $H$, is reduced to a reversal field, $H_R$. The magnetization, $M$, is measured as $H$ is increased from $H_R$ to positive saturation. This is performed for a series of $H_R$, collecting a family of FORCs. A mixed second order derivative extracts the normalized FORC distribution, $\rho(H, H_R) = -\frac{1}{2M_S}\frac{\partial^2 M}{\partial H \partial H_R}$. The dominant magnetic interaction was also probed with $\Delta M$ measurements:[35-37] $\Delta M(H) = \frac{M_{DCD}(H)}{M_R} + 2\frac{M_{IRM}(H)}{M_R} - 1$, where $M_{DCD}(H)$, $M_{IRM}(H)$ and $M_R$ are the dc demagnetization remanence, isothermal remanence, and saturation remanence, respectively. The value of $\Delta M$ will be positive for magnetizing (exchange) interactions, and negative for demagnetizing (dipolar) interactions.[17,35,37] Atomic and magnetic force microscopy (AFM/MFM) were performed using a low-moment CoCr coated Si cantilever, in phase-detection mode.

Polarized neutron reflectometry was performed at the NIST Center for Neutron Research on the polarized beam reflectometer (PBR) using 4.75 Å wavelength neutrons.[44,45] Measurements were performed with an out-of-plane guide field. In this geometry, the non-spin-flip scattering is sensitive to the depth profile of the nuclear scattering length density ($\rho_N$, indicative of the nuclear composition), while the spin-flip scattering is sensitive to the depth profile of the magnetic scattering length density ($\rho_M$, proportional to the magnetization).[46] Modeling of the data was performed using the Refl1D software package.[47] Parallel fitting of datasets with coupled parameters was performed to promote uniqueness of the fit.

**Micromagnetic simulations**. Three dimensional micromagnetic simulations were performed using the object oriented micromagnetic framework (OOMMF) platform.[21] For a vertical construction of $L1_0$-FePt($t_{L10}$)/$A1$-FePt($t_{A1}$) with $t_{L10}$ = 4 nm and variable $t_{A1}$ (1 nm, 5 nm, and 9 nm), we simulated blocks with 100 nm × 100 nm × ($t_{L10}+t_{A1}$) and 1000 nm × 10 nm × ($t_{L10}+t_{A1}$), using a 1 nm cubic mesh, and a larger block of 1500 nm × 1500 nm × 10 nm, using a 6 nm × 6 nm × 1 nm rectangular mesh.[21] The $L1_0$-FePt layer was modeled with saturation magnetization $M_S = 1150$ emu·cm$^{-3}$ (1 emu = $10^{-3}$ A·m$^2$), anisotropy $K_U = 3.6$ J·cm$^{-3}$, exchange constant $A = 1.3 \times 10^{-13}$ J·cm$^{-1}$ and the polycrystalline $A1$-FePt layer as $M_S = 1150$ emu·cm$^{-3}$, $K_U = 0$, $A = 1.3 \times 10^{-13}$ J·cm$^{-1}$.[48]

**Acknowledgements**

This work has been was supported by the NSF (DMR-1008791 and DMR-1543582), BaCaTec (A4 [2012-2]), the France-Berkeley Fund, and DFG (Wi 1828/4-2). Work at NTHU supported by the Ministry of Science and Technology of Republic of China under Grant No. 104-2221-E-007-047-MY2 and MOST 104-2221-E-007-046-MY2. D.A.G. acknowledges support from the National Research Council Research Associateship Program.

**Author contributions**

D.A.G. and K.L. conceived and designed the experiments. J.W.L. and C.H.L. synthesized the samples. D.A.G. and K.L. performed magnetic characterizations. D.A.G. and B.J.K. carried out PNR investigations. M.W. and D.A.G. did micromagnetic simulations. D.A.G. and K.L. wrote the first draft of the manuscript. K.L. coordinated the project. All authors contributed to discussions and manuscript revision.

**Additional information**

Supplementary information is available in the online version of the paper. Reprints and permissions information is available online at www.nature.com/reprints.

**Competing financial interests**

The authors declare no competing financial interests.

**Figure Captions:**

**Figure 1. Magnetometry.** Major hysteresis loops of $L1_0$-FePt (4 nm)/$A1$-FePt ($t_{A1}$) bilayer films in the (**a**) perpendicular and (**b**) in-plane orientation, normalized to the saturation magnetic moment of the $L1_0$-FePt. Samples are identified by color and symbol for $t_{A1}$=0 nm (black squares), 2 nm (red circles), 5 nm (blue triangles), and 9 nm (green inverted triangles).

**Figure 2. Polarized neutron reflectometry.** Polarized neutron reflectivity of $L1_0$-FePt(4 nm)/$A1$-FePt($t_{A1}$) with (**a**) $t_{A1}$=2 nm and (**b**) $t_{A1}$=5 nm, where the non-spin-flip channels are identified as black solid squares ($R^{++}$) and blue solid circles ($R^{--}$), and spin-flip ones as red open triangles ($R^{+-}$) and green open inverted triangles ($R^{-+}$). Schematic insets show the spin configurations in the $L1_0$ layer (red) and $A1$ layer (purple), respectively. The corresponding nuclear and magnetic scattering length density (SLD) depth profiles for the samples with $t_{A1}$ of 2 nm and 5nm are shown in panels (**c**) and (**d**), respectively, which are also used to calculate the reflectivity shown in (**a**, **b**) as solid lines. The nuclear depth profile is shown as a black line, and the magnetic profile as a red line; background colors from left to right identify the $SiO_2$ (grey), $L1_0$-FePt (red), $A1$-FePt (blue), Ti Cap (green), and air (white).

**Figure 3. FORC distribution and $\Delta M$ measurements.** (**a-d**) FORC distributions and (**e**) $\Delta M$ results of $L1_0$-FePt (4 nm) /$A1$-FePt($t_{A1}$) films where $t_{A1}$ is (**a**) 0 nm, (**b**) 2 nm, (**c**) 5 nm, and (**d**) 9 nm. Positive peaks of $\Delta M$ are characteristic of magnetizing (e.g. exchange) interactions, while negative ones indicate demagnetizing (e.g. dipolar) interactions. In panel (**e**) samples are identified by color and symbol for $t_{A1}$=0 nm (black squares), 2 nm (red circles), 5 nm (blue triangles), and 9 nm (green inverted triangles).

**Figure 4. Magnetic domain structure.** MFM images at remanence after ac demagnetization of $L1_0$-FePt/ $A1$-FePt($t_{A1}$) with $t_{A1}$ of (**a**) 0 nm, (**b**) 2 nm, (**c**) 5 nm, and (**d**) 9 nm. The scale bar indicates 2μm.

**Figure 5. Micromagnetic simulations.** (**a**) Cross-section through a domain-wall at remanence after in-plane saturation for a block with dimensions $x=100$ nm, $y=100$ nm, $z=(t_{L10}+t_{A1})$, where $t_{L10} = 4$ nm and $t_{A1} = 2$ nm. Top and bottom panels show out-of-plane and in-plane magnetization, respectively, across a domain wall for selected layers ($z=1$ nm and 4 nm, i.e., base and top of the $L1_0$-FePt, and 5 nm and 6 nm, i.e., base and top of the $A1$- FePt). Dashed curves show theoretical magnetization curves, vertical dashed lines represent the theoretical domain-wall width $\delta_{DW} \propto \sqrt{A/K}$. (**b**) as in (**a**), but now for $t_{A1} = 9$ nm. (**c**) Contributions to free energy density $F$ (top panel) along the magnetization curve shown in the bottom panel. All energy terms are normalized to the stray field energy constant. (**d**) Side-view example magnetization configuration in a long narrow strip ($x=1000$ nm, $y=10$ nm, $z=(t_{L10}+t_{A1})$, with $t_{L10} = 4$ nm and $t_{A1} = 5$ nm) hosting domain walls in both the $L1_0$ and $A1$ layer, under in-plane field ($H_x = -0.1$ T). Color coding indicates $x$-component of the magnetization. (**e, f**) Example magnetic configuration from large model ($x=1500$ nm, $y=1500$ nm, $z=10$ nm) after domain nucleation, with color contrast in (**e**) indicating the out-of-plane magnetization, and in (**f**) indicating the into-the-page magnetization.

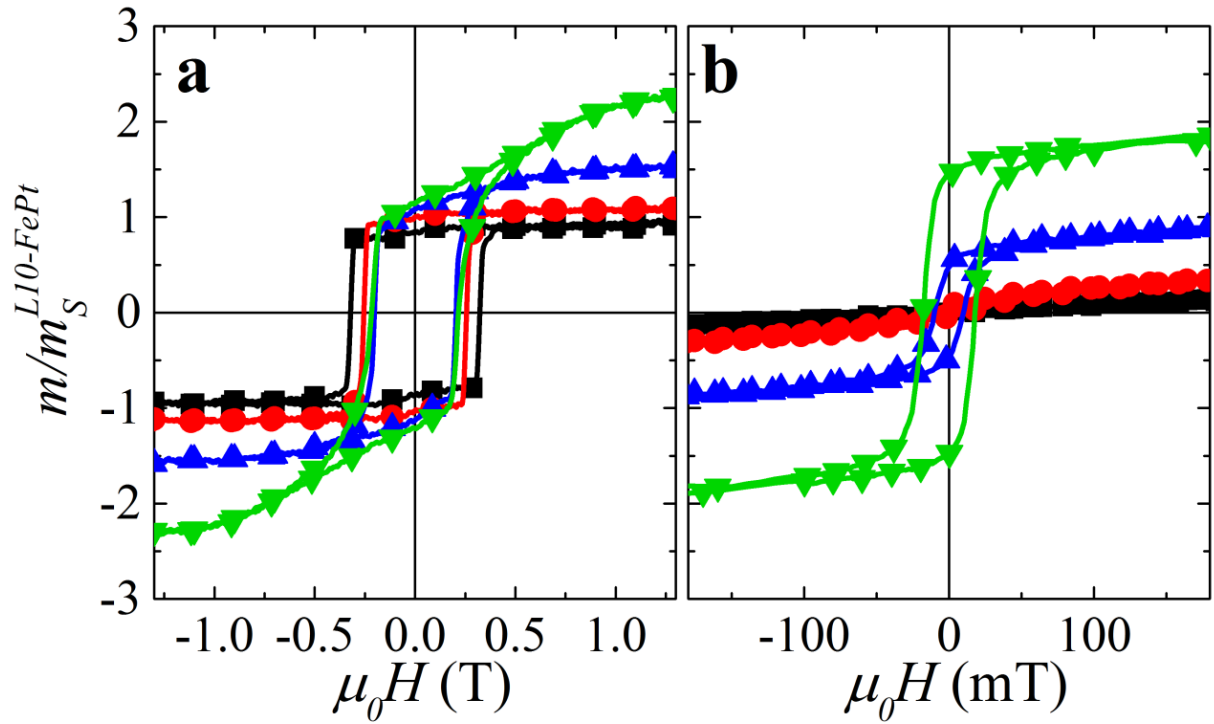

**Fig. 1**

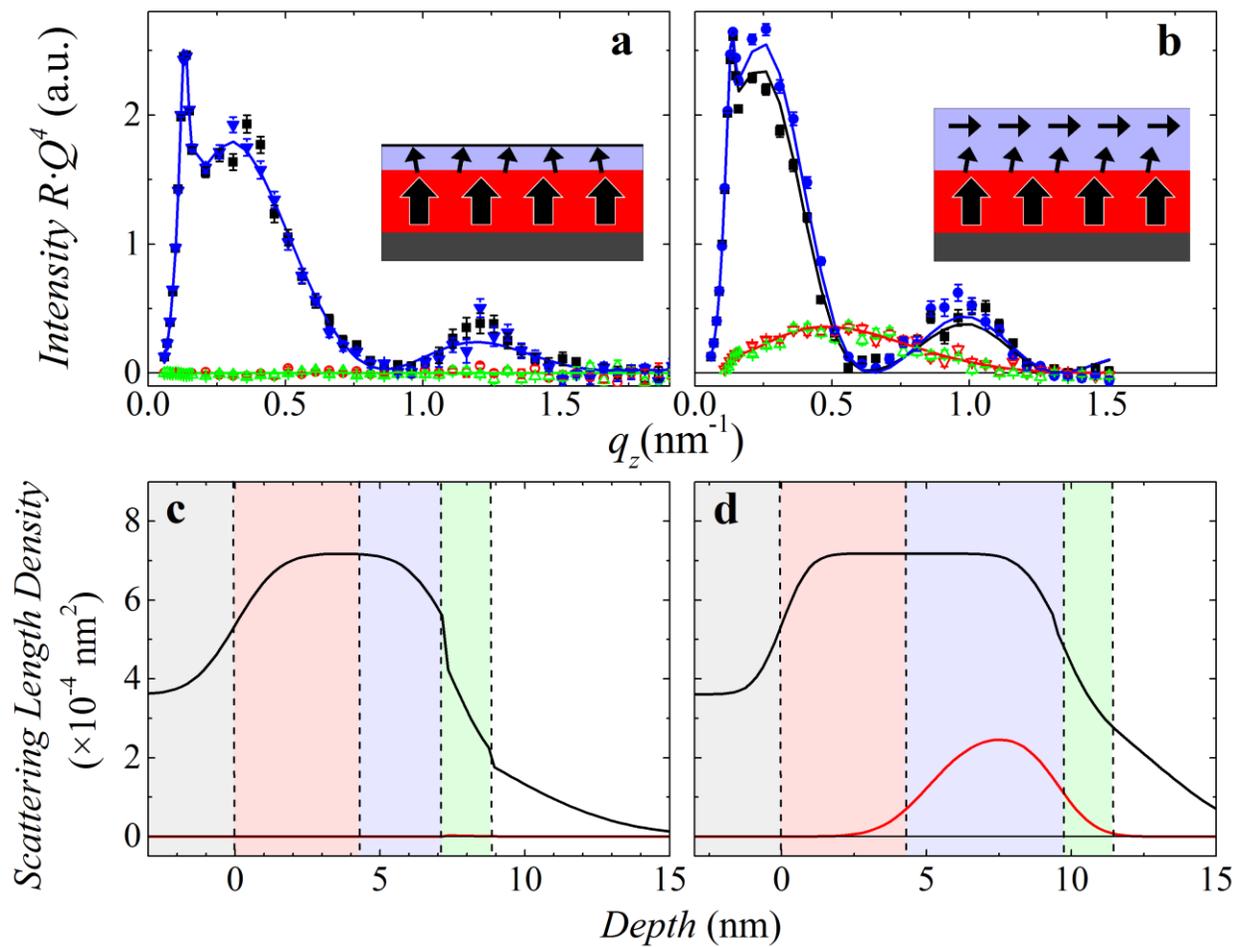

**Fig. 2**

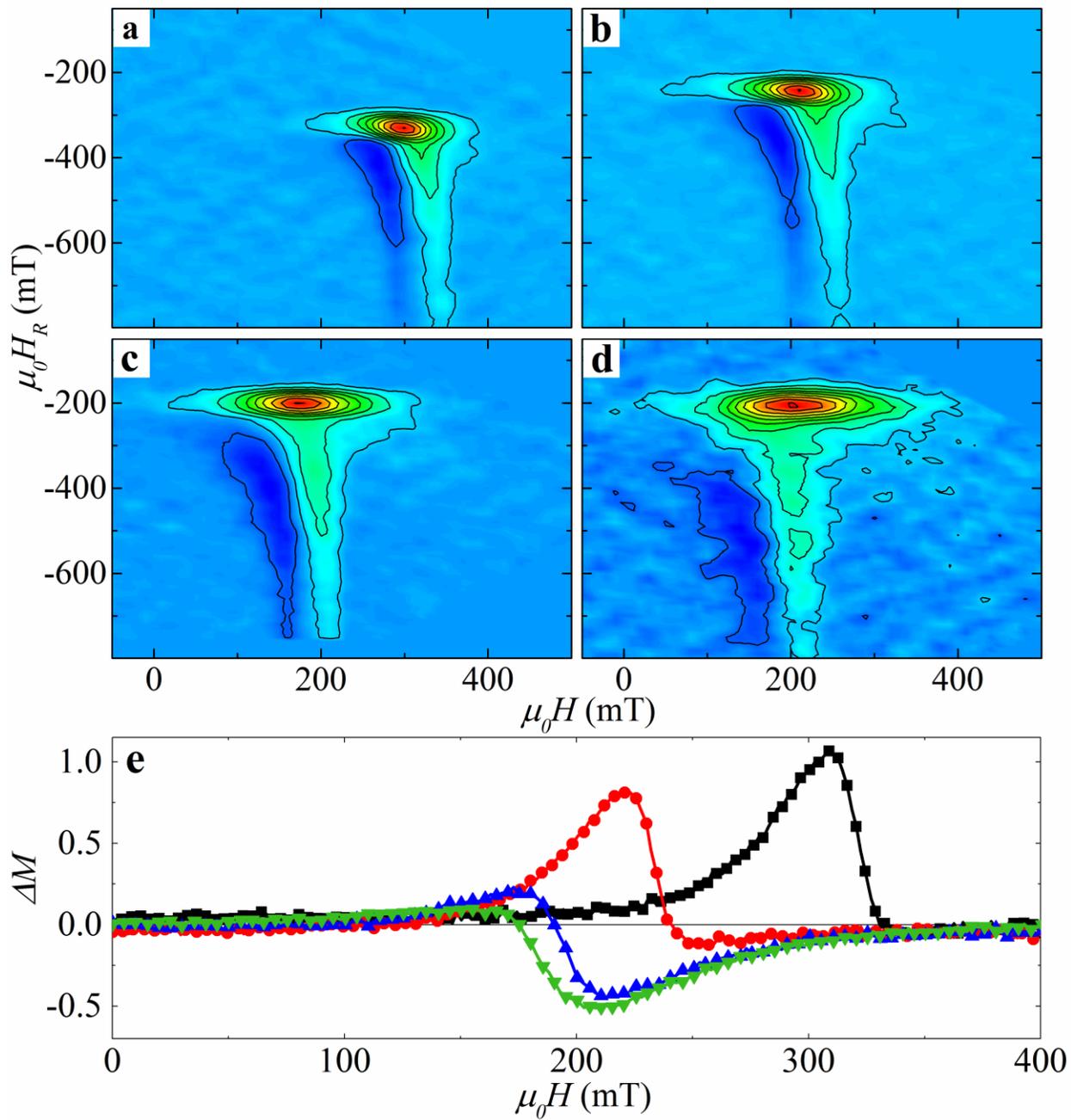

**Fig. 3**

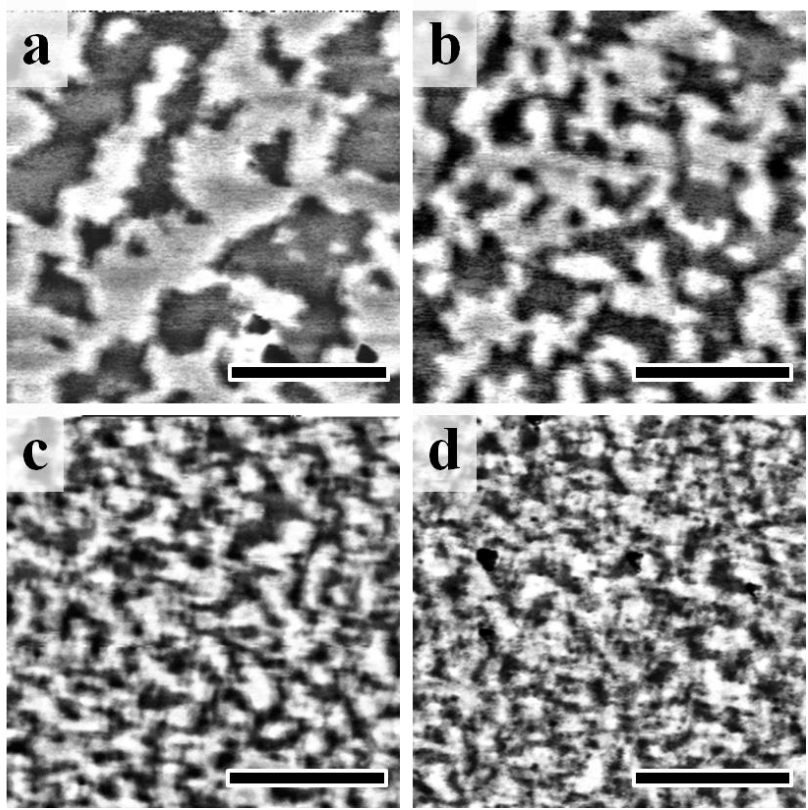

**Fig. 4**

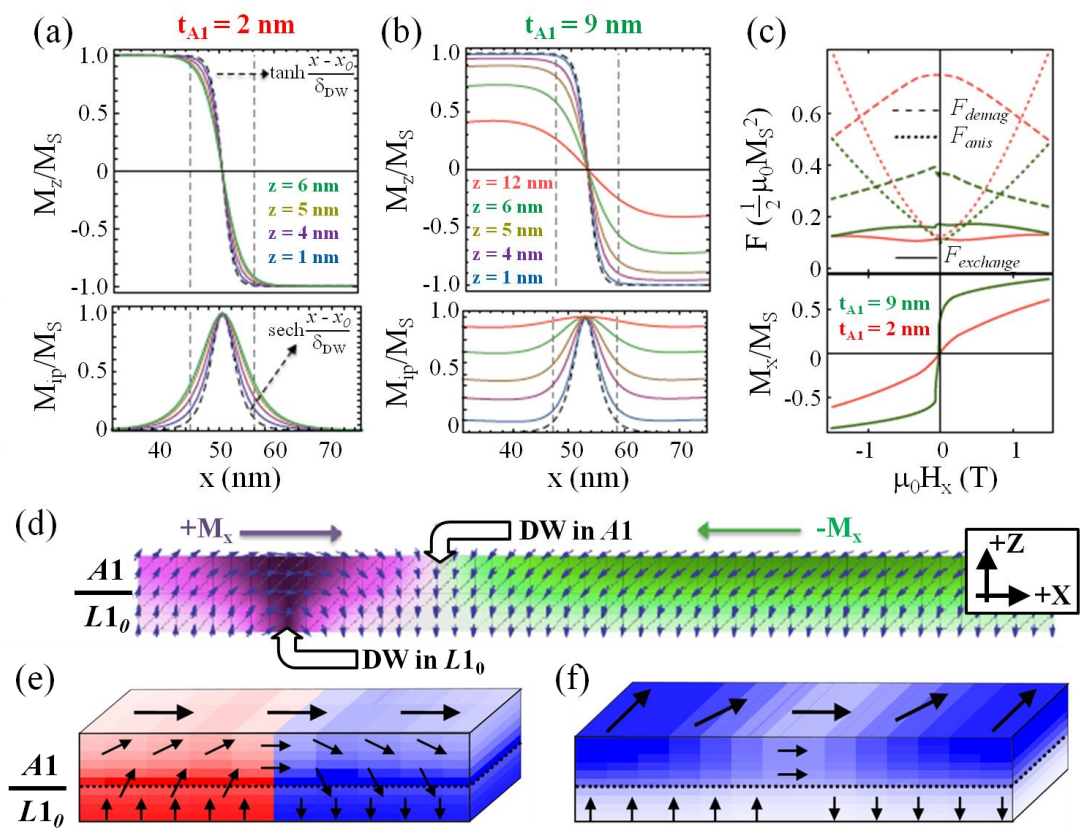

**Fig. 5**

# Magnetic Yoking and Tunable Interactions in FePt-Based Hard/Soft Bilayers

Dustin A. Gilbert,[1,2] Jung-Wei Liao,[3] Brian J. Kirby,[2] Michael Winklhofer,[4,5,6]
Chih-Huang Lai,[3] and Kai Liu[1,*]

[1]Dept. of Physics, University of California, Davis, California 95616, USA
[2]NIST Center for Neutron Research, Gaithersburg, Maryland 20899, USA
[3]Dept. of Materials Science and Engineering, National Tsing Hua University, Hsinchu 30013, Taiwan
[4]Dept. of Earth and Environmental Sciences, Geophysics, Munich University, 80333 Germany
[5]Faculty of Physics, University of Duisburg-Essen, 47057 Duisburg, Germany
[6]IBU, School of Mathematics and Science, Carl von Ossietzky University, 26129, Oldenburg, Germany

**Supplementary Information**

Similar experiments were performed on $L1_0$-FePt/Fe bilayer films. The major loop behavior, Supplementary Fig. S1, is similar to those with $A1$-FePt films featured in the main text. Comparing the relative changes of the $A1$-FePt and Fe soft layers, the Fe reduces the bilayer coercivity much more significantly (from 340 mT to 165 mT for 2nm Fe vs. 252 mT for 2 nm $A1$-FePt) than the $A1$-FePt. The FORC distributions for these films are shown in Supplementary Fig. S2. Similar to the distributions in the main text, the FORCs consist of a vertical and horizontal ridge features. Here, even the thinnest Fe layer causes the FORC distribution to shift to a 'T' configuration, and evolve continuously to a right bending construction with thicker Fe. $\Delta M$ measurements, shown in Supplementary Fig. S3, confirm that for all thicknesses of Fe the dipolar interactions are dominant. For the Fe soft layer the dipolar interactions are enhanced more than in the $A1$-FePt case likely due to an enhanced yoking effect from the higher $M_S$ in Fe (1700 emu·cm$^{-3}$, 1 emu = 10$^{-3}$ A·m$^2$) compared to $A1$-FePt (1200 emu·cm$^{-3}$). An increased $M_S$ also favors an in-plane magnetization at thinner soft-layer thicknesses due to the gain in magnetostatic energy.

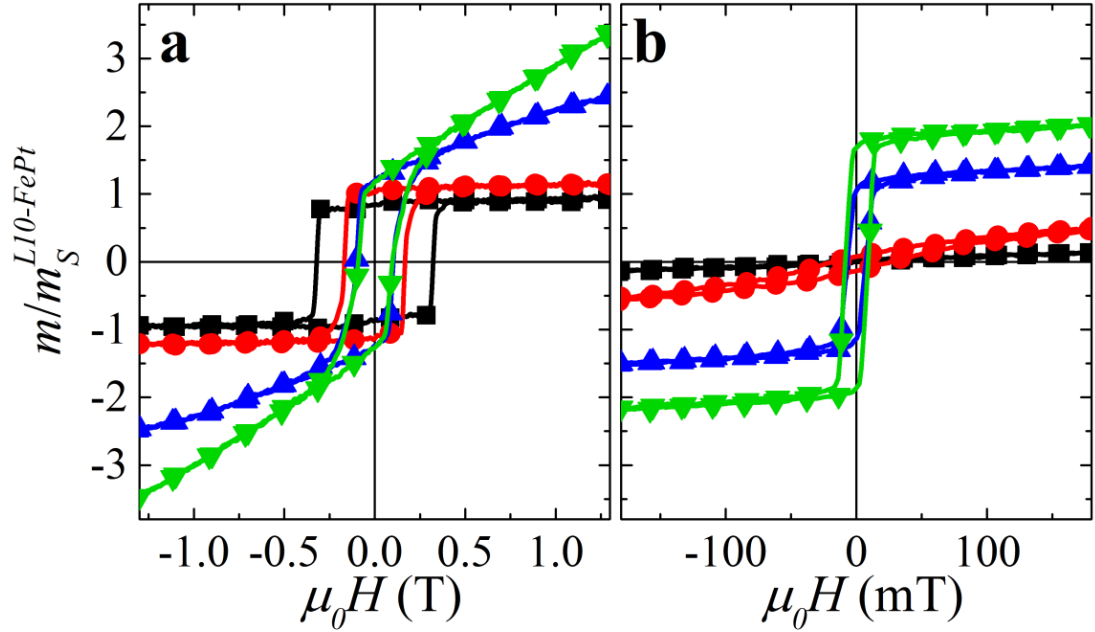

**Supplementary Fig. S1. Magnetometry.** Major Hysteresis Loops of $L1_0$-FePt (4 nm) / Fe ($t_{Fe}$) films with the magnetic field applied in the (**a**) perpendicular and (**b**) in-plane orientation. Samples are identified by color and symbol for $t_{Fe}$=0 nm (black squares), $t_{Fe}$=3 nm (red circles), $t_{Fe}$=5 nm (blue triangles), and $t_{Fe}$=9 nm (green inverted triangles).

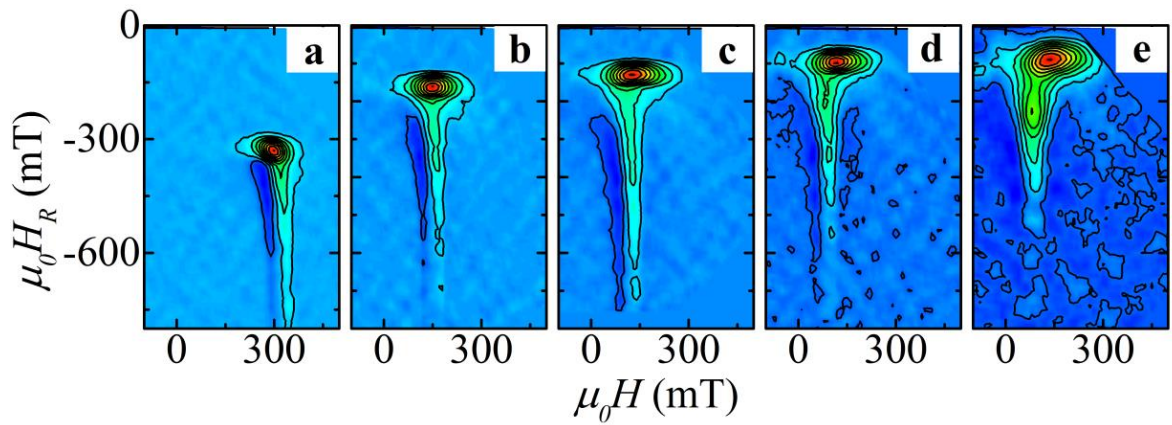

**Supplementary Fig. S2. FORC distributions.** FORC distributions of $L1_0$-FePt/Fe($t_{Fe}$) films where $t_{Fe}$ is (**a**) 0 nm, (**b**) 2 nm, (**c**) 3 nm, (**d**) 5 nm, and (**e**) 9 nm.

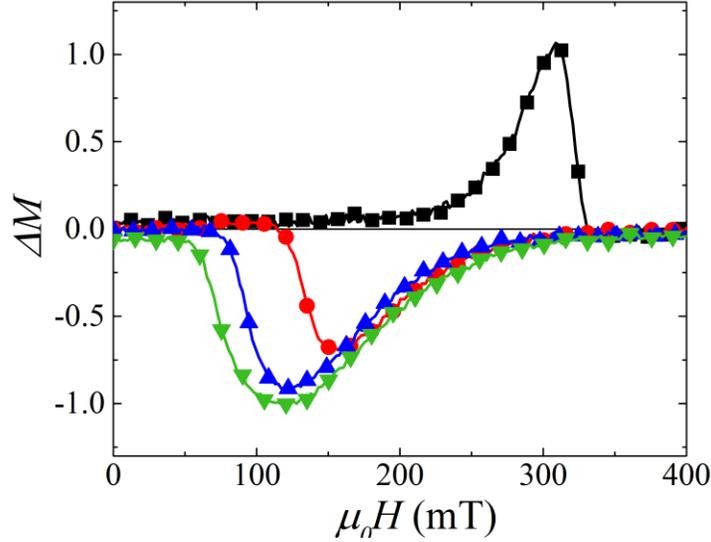

**Supplementary Fig. S3. *ΔM* measurements.** *ΔM* plot for $L1_0$-FePt (4 nm) /Fe($t_{Fe}$) films. Samples are identified by color and symbol for $t_{Fe}$=0 nm (black squares), $t_{Fe}$=3 nm (red circles), $t_{Fe}$=5 nm (blue triangles), and $t_{Fe}$=9 nm (green inverted triangles).

To estimate the strength of the dipolar field in the $L1_0$-FePt/$A1$-FePt system, we have modeled the system as 280 nm sized square domains with opposite out-of-plane orientations (found in bare $L1_0$-FePt film) with a 5 nm non-contributing domain wall region. The in-plane dipolar field from these domains was calculated at 4 nm above the $L1_0$-FePt film. The net in-plane dipolar field is found to be greater than 100 mT over the domain wall, and greater than 4 mT - the $A1$-FePt coercivity - up to 50 nm (laterally) from the domain boundary. Thus, it is reasonable that the soft layer orientation indeed follows the dipolar fields (yoking) and as a result acts to moderate interactions between domains in the hard layer.